\documentclass[useAMS,usenatbib]{mn2e}
\usepackage{graphicx}
\usepackage{times}

\newcommand{\SiII}{\hbox{{\rm Si}{\sc \,ii}}}

\newcommand{\ZnII}{\hbox{{\rm Zn}{\sc \,ii}}}

\newcommand{\MgII}{\hbox{{\rm Mg}{\sc \,ii}}}

\newcommand{\HI}{\hbox{{\rm H}{\sc \,i}}}

\newcommand{\lya}{\hbox{{\rm Ly}$\alpha$}}

\newcommand{\cmsq}{\hbox{cm$^{-2}$}}
\newcommand{\NHI}{\hbox{$N_{\HI}$}}

\newcommand{\EW}{\hbox{$W_{\rm r}^{\lambda2796}$}}
\newcommand{\Z}{\hbox{[X/H]}}

\def\bsp_small{\vspace{0.5cm}\small\noindent This paper
has been typeset from a \TeX / \LaTeX\ file prepared by the author.}

\title[Bimodality of DLAs]{Towards Understanding  the Mass-Metallicity relation of Quasar Absorbers:
Evidence for bimodality and consequences.}
  \author[N. Bouch\'e]{Nicolas
  Bouch\'e \\
   $^1$Max Planck Institut f\"ur extraterrestrische Physik,
Giessenbachstrasse, Garching D-85748, Germany; nbouche@mpe.mpg.de 
}

\begin{document}

\date{Accepted 2008 June 7.  Received 2008 May 29; in original form 2008 April 7}

\pagerange{\pageref{firstpage}--\pageref{lastpage}}
%\pubyear{2005}

\maketitle

\label{firstpage}

%%%%%%%%%%%%%%%%%%%%   ABSTRACT  %%%%%%%%%%%%%%%%%%%%%%%

\begin{abstract} 
One way to characterize and understand \HI-selected galaxies is to study their metallicity
properties. In particular, we show that the metallicity of absorbers is a bivariate function of
the \HI\ column density (\NHI) and the \MgII\ equivalent width (\EW).
Thus, a selection upon \EW\  is not equivalent to a \HI\ selection for intervening absorbers.
A direct consequence  for damped absorbers with $\log \NHI>20.3$ that falls  
from the bivariate metallicity distribution is that  any correlation between the  metallicity
\Z\ and velocity width (using \EW\ as a proxy) cannot be interpreted 
as a signature of the mass-metallicity relation akin to normal field galaxies.
In other words,  DLA samples are intrinsically heterogeneous and
the \Z--\EW\ or \Z--$\Delta v$ correlation reported in the literature arises
from the \HI\ cut.
On the other hand, a sample of \MgII-selected absorbers, which are  
statistically dominated by lowest \NHI\ systems (sub-DLAs) at each \EW,   
are found to have a more uniform metallicity distribution.
We postulate that the bivariate distribution [\Z(\NHI,\EW)] can be explained
by two different physical origins of absorbers, namely sight-lines through the ISM of small galaxies 
and sight-lines through   out-flowing material.
Several published results follow from the bivariate \Z\ distribution such as
(a) the properties of the two classes of DLAs, reported by Wolfe et al.,   and
(b) the constant dust-to-gas ratio for \MgII-absorbers.
%Finally, our results show that the mass-velocity width anti-correlation  of Bouch\'e et al. (2006) 
%is can be understood the same context.
\end{abstract}


%%%%%%%%%%%%%%%%%%%%%%%%% INTRODUCTION %%%%%%%%%%%%%%%%%%%%%%%
\section{Introduction}

Among quasar (QSO) absorption line systems, damped \lya\ absorbers (DLAs) 
with column densities $N_{\HI}\ge2\times10^{20}$~\cmsq\ are the most puzzling. 
Their high column density range indicates that they ought to trace cold neutral gas, 
hence disks \citep[e.g.][]{WolfeA_86a,WolfeA_98a}.
This is further supported as in the local universe most of the \HI\ gas with these column densities are in galaxies
\citep[e.g.][]{RosenbergJ_03a,ZwaanM_05a}.
In addition, \citet{ProchaskaJ_98a} (and others) argued that the kinematics
of DLAs (traced by low-ions) are best explained by models
of rapidly rotating disks.

Despite decades of studies,
it is not clear whether this cold neutral gas is part of the interstellar medium of large spirals \citep{WolfeA_86a},
part of the halos of galaxies \citep{BahcallJ_69a} 
or part of dwarf galaxies \citep{YorkD_86a} akin to  the Magellanic Clouds. An alternative
scenario for absorption-selected galaxies
 is that the gas seen in absorption is  part of cold gas clumps in the host galaxy halos
 entrained in outflows produced by supernovae (SNe)  \citep{NulsenP_98a,SchayeJ_01a}.

Among the many different approaches that have been used to make further progress on this issue, one  
is to use the metallicity of the gas. Since the metallicity of the absorbing gas,
and of  DLAs in particular, is typically 1/30$th$ solar \citep[e.g.][]{PettiniM_99a,ProchaskaJ_99b,PerouxC_03b,ProchaskaJ_03d}, and 
 $z\gg 1$ star-forming galaxies have metallicities closer to solar, i.e.
more metal rich by a factor of 10 \citep[e.g.][]{ErbD_06a}, one is often forced to invoke
metallicity gradients in galaxies to 30~kpc and beyond to reconcile the two observations.
If indeed absorption and emission metallicities are related to one another solely by
a metallicity gradient, one would expect a similar mass-metallicity correlation in
DLAs as observed in normal star-forming galaxies \citep{TremontiC_04a,ErbD_06a}.

While  the metallicity of absorbers is usually straightforward to determine
\citep[e.g.][]{ProchaskaJ_00d,ProchaskaJ_03d,PerouxC_03b,PerouxC_06a,ProchaskaJ_06a,KulkarniV_07a}, the mass (dynamical or baryonic)
 of absorption-selected galaxies
cannot be usually measured directly because  the host-galaxy is usually elusive.
Progress is underway from (i) direct dynamical mass estimates of 14 \MgII-selected galaxies
by \citet{BoucheN_07a} (and Bouch\'e, Murphy \& P\'eroux, in prep.),
 and (ii) direct comparison between the host-emission metallicity
and the absorption metallicity (Bouch\'e et al. 2008, in prep.).

Indirect, and statistical, mass measurements have been made from clustering analysis \citep{BoucheN_06c}
  for strong $z\simeq0.5$ \MgII\ absorbers  
with rest-frame equivalent widths $\EW\ge1$\,\AA, which  
are intimately related to low-$z$ DLAs \citep{ChurchillC_00b,RaoS_00a,RaoS_06a}.
\citet{BoucheN_06c} found that the host halo-mass ($M_h$) and 
  \EW\ are anti-correlated. 
 For virialized clouds,  the host  mass and the line-of-sight  velocity dispersion 
 ought to be correlated, i.e.
 a  $M_h$--\EW\ correlation is expected  because  \EW\ is 
  a measure of  the line-of-sight velocity width ($\Delta v$)
  as individual \MgII\ absorptions are saturated \citep{EllisonS_06a}.   
Thus, the results of \citet{BoucheN_06c} imply that
\MgII\ clouds are not virialized in the host halos, and super-novae driven
  outflows provide a natural mechanism. While the mass-\EW\ results have been confirmed
  by an independent team (Allen, Hewett \&\ Ryan-Weber 2008, in prep.), 
%  the outflow interpretation 
  ad-hoc models have been proposed to explain the anti-correlation in a cosmological context
  \citep[e.g.][]{TinkerJ_08a}.

Since one would   expect a mass--metallicity correlation for all galaxies, 
the mass-\EW\ anti-correlation of \citet{BoucheN_06a} implies a
metallicity-\EW\ anti-correlation, or equivalently  a metallicity-velocity width anti-correlation.
However, several groups have reported just the opposite:  
the metallicity in DLAs correlates  either with the velocity width $\Delta v$
\citep{WolfeA_98a,PerouxC_03b,LedouxC_06a,ProchaskaJ_08a}
or with the \EW\ \citep{MeiringJ_06a,MurphyM_07a}. 
It is tempting to assume that $\Delta v$
is a measure of the line-of-sight velocity dispersion, i.e. that it correlates with the mass of the host-galaxy,
since when combined with the above results, one would naturally imply a normal mass-metallicity relation.

Thus, there appears to be a conflict between the $M_h$-\EW\ anti-correlation of \citet{BoucheN_06c}
and the \EW(velocity)--\Z\ correlations reported in the literature.
In this paper, we show that the conflict is apparent and reflects the various selections at play
(\HI\ vs. \MgII) using \MgII\ systems from the literature.
In section~2, we describe our sample
where we collected neutral column density \NHI, \MgII\ equivalent widths
 and metallicities \Z. Section~3 shows our results.

\section{Data}

%Strong \MgII\ absorbers are closely related to DLAs. In particular,
%at $z\simeq0.5$, 
%the DLA fraction is essentially 0 below $\EW\la 0.6$\,\AA,
%then  is about $\sim35$\%\ for all \MgII absorbers with $\EW\ga 0.6$\,\AA\ \citep[e.g.][]{RaoS_06a}
%and may reach about $\ga50$\%
%for \MgII\ absorbers with $\EW\simeq 3$~\AA.

We combined various samples of \MgII\ absorbers and DLAs from the literature,
namely we used the samples of \citet{RaoS_06a},
\citet{EllisonS_06a}, \citet{KulkarniV_05a}, \citet{CurranS_07a}, \citet{MurphyM_07a}
\citet{LedouxC_06a}, augmented by the catalog of \citet{RyabinkovA_03a}.
The \HI\ column densities come mostly from the STIS survey of \citet{RaoS_00a,RaoS_06a}
for the low-redshift absorbers.
The entire catalog contains about 1200~absorbers, of which 377 have both
\EW\ and \NHI\ measured.
We then match the literature  samples  with published metallicity measurements from \citet{PerouxC_03a},
\citet{PerouxC_04a}, \citet{PerouxC_06a}, 
 \citet{MollerP_04a}. \citet{ProchaskaJ_06a}, \citet{KulkarniV_05a},
 \citet{LedouxC_06a}, \citet{EllisonS_06a}, \citet{ProchaskaJ_06a}, \citet{MeiringJ_06a}, \citet{MeiringJ_08a} and
 \citet{MurphyM_07a}. The final sample is made of 89 absorbers with known \NHI, \EW, and \Z.

\begin{figure}
\centerline{\includegraphics[width=7cm]{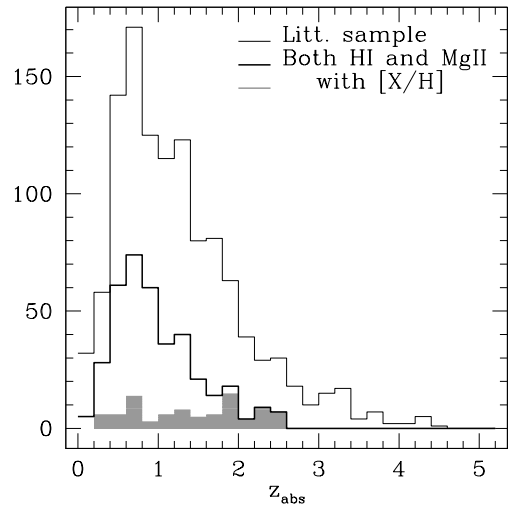}}
\caption{Redshift distributions of the total sample (solid line), of the \HI-\MgII\ sub-sample (thick line)
and of the final sample  (shaded histogram) with metallicity measurements restricted to [Zn/H].
\label{fig:redshift}}
\end{figure}

We show the redshift distribution of the sub-samples in Fig.~\ref{fig:redshift}.
The solid histogram shows the literature sample of 1200 absorbers. The thick histogram
shows the 377 absorbers with \EW\ and \NHI, and the grey histogram
shows the 89 absorbers with \EW, \NHI\ and [Zn/H].
In order to have  homogeneous metallicity measurements, we impose
that all metallicity measurements are from Zn.
In order to probe for any redshift evolution, we will split the sample
into low-$z$ ($z<1.6$) and high-$z$ ($z>1.6$).

\section{Results}

\begin{figure*}
\centerline{\includegraphics[width=7cm]{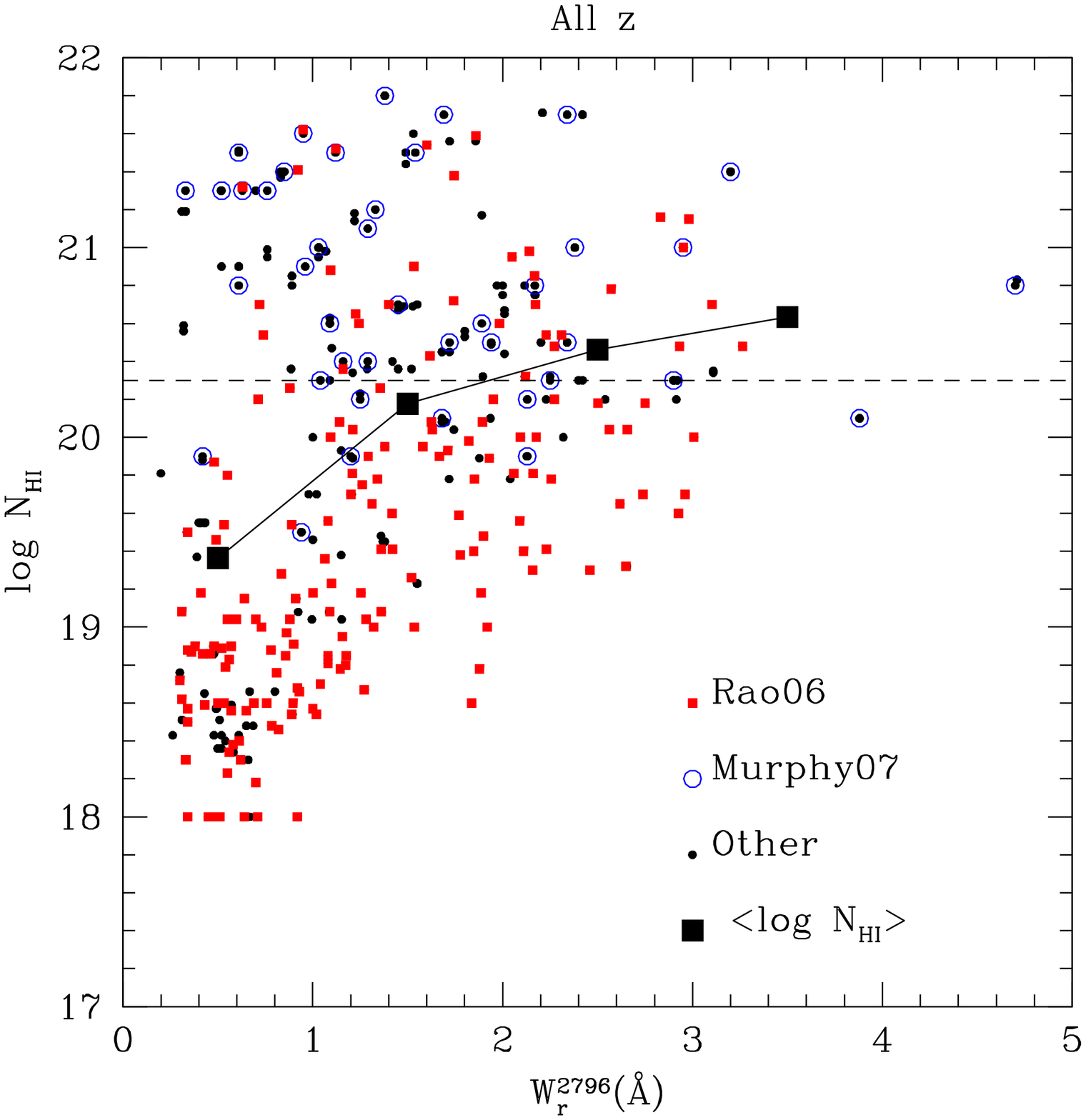}
\includegraphics[width=7cm]{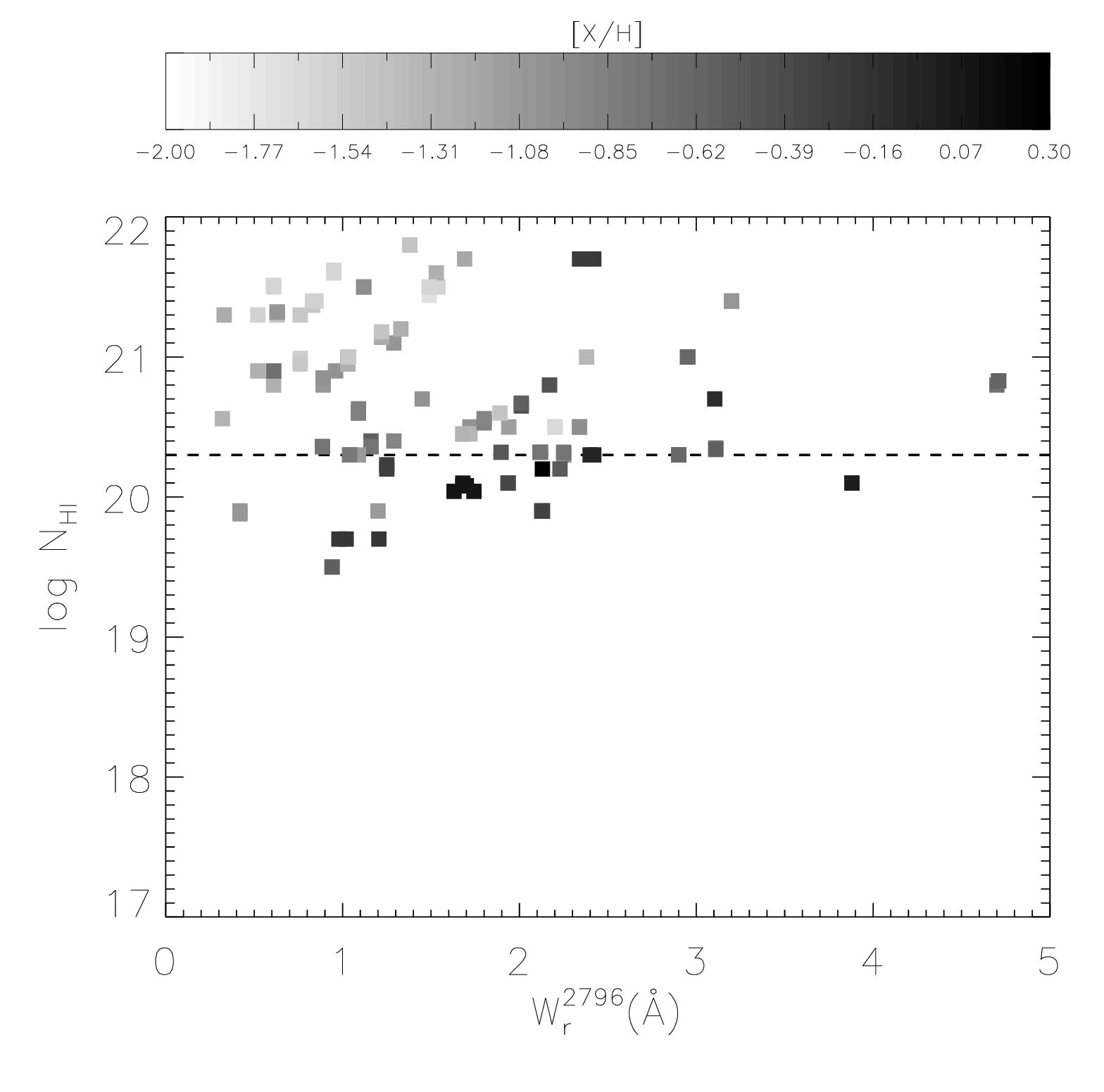}}
\caption{{\bf a)} The distribution of absorbers in the \NHI-\EW\ plane.
Strong \MgII\ absorbers have indeed larger \NHI, and can be regarded as DLA-dominated,
however the opposite is not true. A DLA sample covers the entire range of \EW.
The large  connected squares show that the logarithmic mean $<\log \NHI>$.
 {\bf b)} For those with [Zn/H] measurements, the points are color-coded according
 to increasing metallicity from [Zn/H]$=$-2 to 0.
There is a clear metallicity gradient. Furthermore, the gradient is inclined with respect to the $y$-axis.
In both panels, the dashed line shows the classical DLA threshold of 20.3.
\label{fig:intro}}
\end{figure*}

\subsection{Metallicity gradient in \NHI-\EW.}

Fig.~\ref{fig:intro}(a) shows the  distribution of absorbers in the \NHI-\EW\ plane 
as in \citet{RaoS_06a}.
Note that strong \MgII\ absorbers have indeed larger \NHI, and can be regarded as DLA-dominated,
however the opposite is not true. A DLA sample covers the entire range of \EW.
The large  connected squares show  the logarithmic mean $<\log \NHI>$.  
The logarithmic mean is a better statistic to quantify the distribution of the points.
 \citet{RaoS_06a} elected
to use the mean $<\NHI>$ statistics since they were interested in the mean \HI\ column density
in order to constrain $\Omega_{\HI}$.
The solid squares show that $<\log \NHI>$ increases with equivalent width,
reflecting an increasing fraction of DLAs as a function of \EW.

In Fig.~\ref{fig:intro}(a), as noted many times \citep[e.g.][]{RaoS_06a,CheloucheD_07a},
there  are no absorbers to the bottom right of the figure, i.e.
 with large \EW\ and low \HI\ column densities. 
The lack of objects in that part of the diagram is not due to selection effects; strong
\MgII\ absorbers are easy to identify there.
% This can now be explained as ...........

\begin{figure*}
\centerline{ 
\includegraphics[width=7cm]{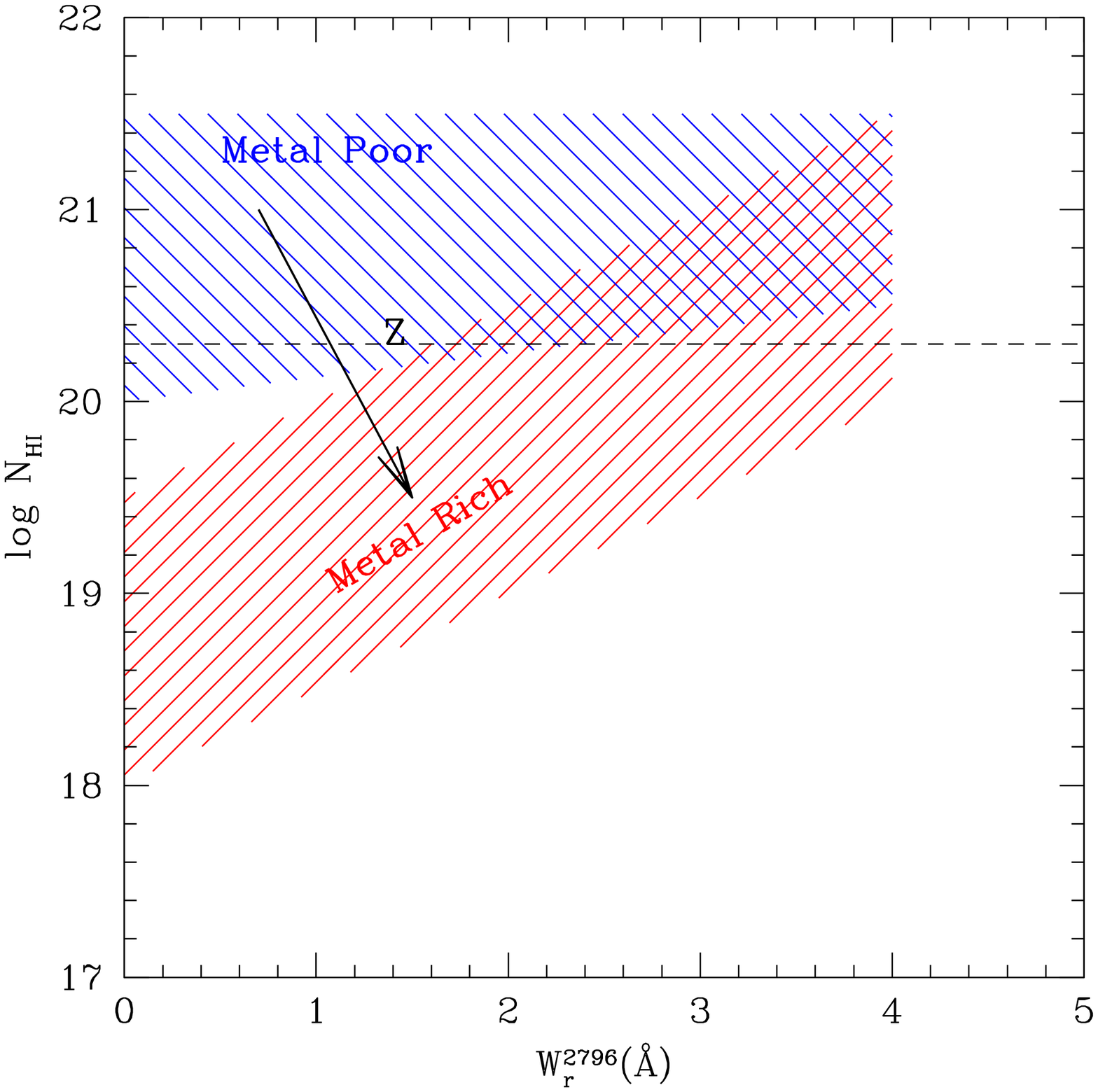} 
\includegraphics[width=7cm]{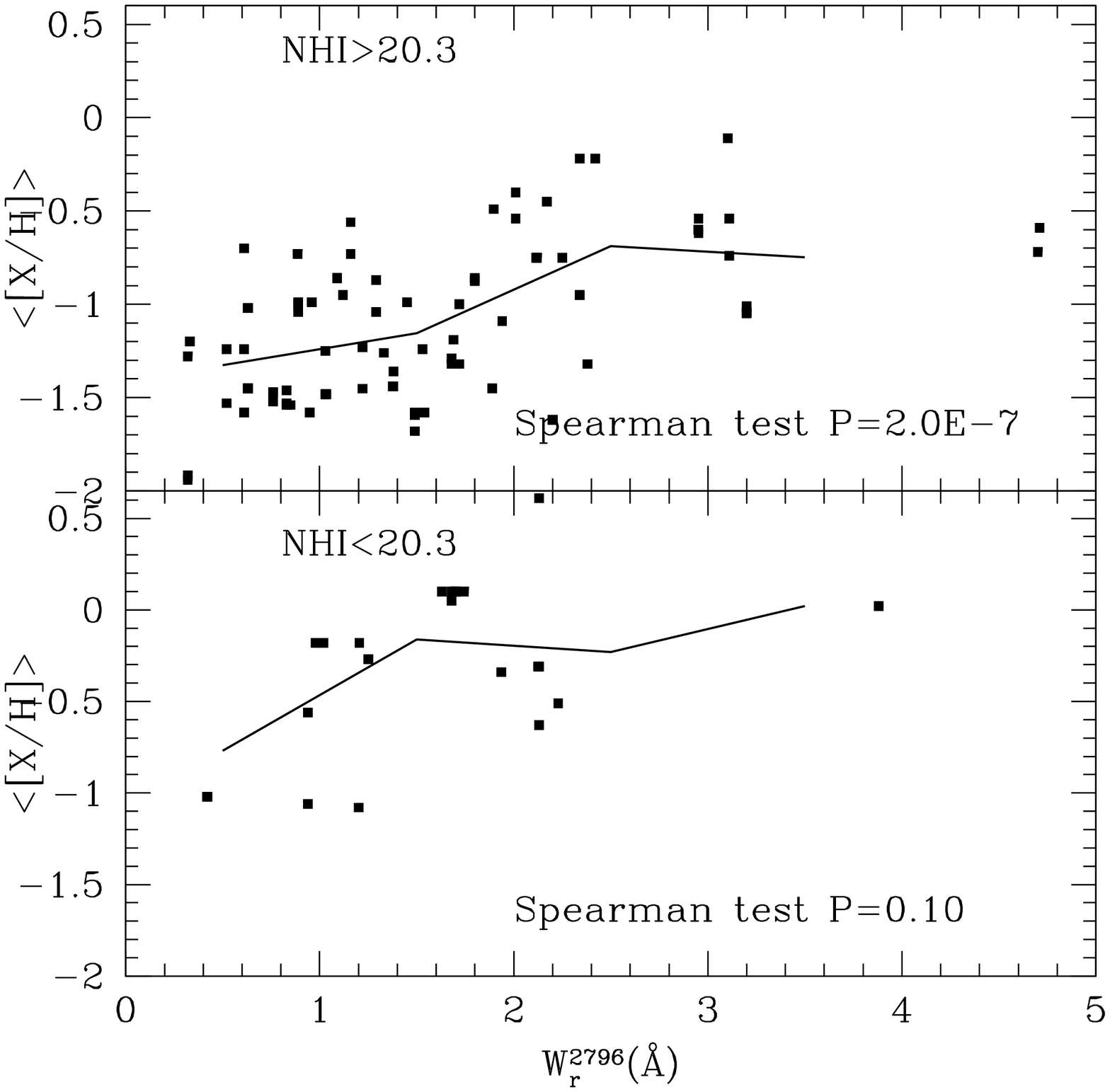} }
\caption{{\bf a)}: Schematic representation of the metallicity gradient shown in Fig.~\ref{fig:intro}(b).
The  `metal poor'
extends over the high $\log \NHI$  region,
while the `metal rich' region 
extends over the low $\log \NHI$  region (assumed to have uniform metallicity).
{\bf b)}: Mean metallicity $\langle\Z\rangle$
 for DLAs with $\log \NHI>20.3$ (top) and sub-DLAs (bottom). 
 The solid lines show the average $\langle\Z\rangle$
 in bins of \EW\ as in Fig.~2(a). 
The increase in the mean metallicity  for DLAs is $\sim0.8$--1~dex which is the increase reported
by \citet{LedouxC_06a}. On the other hand, sub-DLAs have roughly a similar metallicity.
The $P$-values for the Spearman's rank correlation test are shown.
This figure shows that the \Z-$\Delta v$ and \Z-\EW\  correlations reported by  \citet{LedouxC_06a}  for DLAs 
and \citet{MurphyM_07a} are both a `selection effect': the correlation with metallicity originates
from the increased overlap of the metal poor and metal rich systems 
in the \NHI-\EW\ plane (Fig.~\ref{fig:intro}b).
\label{fig:gradient}}
\end{figure*}

The solid squares in Fig.~\ref{fig:intro}(b) show the absorbers whose \Z\ measurement
exist in the literature, and color-coded according to [Zn/H] from -2 to 0.
This figure clearly shows that the more metal poor (lighter grey points) are located in 
a different location as the metal rich (darker grey points).
In other words, there is a strong metallicity gradient across the \NHI--\EW\ plane,
represented schematically in Fig.~\ref{fig:gradient}(a).
The metallicity gradient is noted by the vector.
For low \HI\ column densities  ($\log \NHI<19.5$), there is a lack of metallicity data due to
difficulties in measuring [X/H] due to an increasing ionization correction \citep{PerouxC_06a,ProchaskaJ_06a}.

\subsection{Biased metallicities}

Fig.~\ref{fig:intro}(b) shows that sub-DLAs are generally more metal rich than DLAs,
 a result noted already by many
\citep[e.g.][]{KhareP_07a}. 
 However, since the metallicity changes in a subtle way in the \NHI--\EW\ plane,
 (Fig.~\ref{fig:gradient}[a]), one would expect to have different mean metallicity $\langle\Z\rangle$
 for different   \NHI-selected samples. 
 This is illustrated in Fig.~\ref{fig:gradient}(b), where we plot $\langle\Z\rangle$
 for DLAs with $\log \NHI>20.3$ (top) and sub-DLAs with $\log \NHI<20.3$ (bottom).
The metallicity \Z\ increases as a function of \EW\ 
(a proxy for the velocity width $\Delta v$) for DLAs.
The  $P$-value of the Spearman's correlation test is $2\times10^{-7}$, i.e.
the correlation is significant at $>4$-$\sigma$.
Moreover, the increase in the mean metallicity \Z\ is $\sim0.8$--1~dex, which is the increase reported
by \citet{LedouxC_06a} for their DLA-sample.
On the other hand,   for  absorbers with $\log \NHI$ less than $20.3$, 
the mean metallicity appears constant (Fig.~\ref{fig:gradient}[b], bottom).
The Spearman's correlation test gives a $P$-value much higher (0.10) and shows 
that the correlation is significant at best at 1.5$\sigma$. 
Without the one data point at \EW=0.5\AA, the $P$-value is  higher still: 0.43 and 
 the \EW\  is not correlated with \Z.
We note that the  \Z-\EW\ relation reported by
 \citet{MurphyM_07a} is explained by the fact that
  their sample is dominated by systems with \HI\ column densities mostly above
  the DLA threshold $\NHI>20.3$ (Fig.~\ref{fig:intro}[a]).
Thus, the \Z-$\Delta v$ and \Z-\EW\  correlations reported by  \citet{LedouxC_06a}  for DLAs 
and \citet{MurphyM_07a} are both a `selection effect': it originates
from the increased overlap of the metal poor and metal rich systems in the \NHI-\EW\ plane.

While the fraction of DLAs increases with \EW, this exercise shows that
the selection upon \EW\  is not equivalent to a \HI\ selection, as the two sample
will have very different properties: the \MgII-selected sample will be biased towards more metal rich absorbers,
while the \HI-selected sample will be more metal poor, with a strong metallicity dependence on \EW.

Note that there seems to be no redshift  bias in our results:
 Fig.~\ref{fig:intro}[b] looks similar for the $z_{\rm abs}<1.6$ and $z_{\rm abs}>1.6$ sub-samples,
 with perhaps an overall shift in metallicities which has been before
for DLAs and sub-DLAs \citep{ProchaskaJ_03b,KulkarniV_07a}.
Based on these results, we turn towards a physical interpretation.

%Second, because the selection upon \EW\ for \HI\ absorbers is not equivalent to a \HI\ selection,
% the $M_h$--\EW\ anti-correlation of \citet{BoucheN_06c} is valid for \EW-selected absorbers while the 
% \EW(or $\Delta v$)-$Z$ correlation of \citet{LedouxC_06a,MurphyM_07a} is valid for \NHI-selected
% absorbers. Thus, the apparent contradiction raised in the introduction resulted from the two
% selections being very different.

\subsection{Interpretation}

%\begin{figure}
%\centerline{ \includegraphics[width=7cm]{figs/plot_DLA_MgII_hist.allz.EW.lt1.5.eps} }
%\caption{For absorbers with $\EW\le 1.5$~\AA\ shown in Fig.~\ref{fig:intro},
%the \HI\ column density distribution appears bimodal.
%\label{fig:HIhist}}
%\end{figure}

\begin{figure}
\centerline{
\includegraphics[width=7cm]{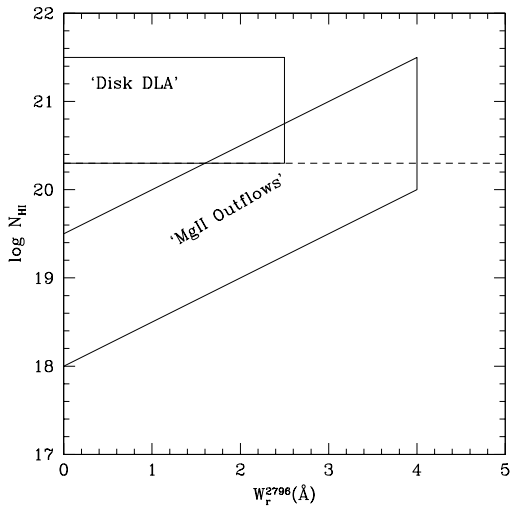}}
\caption{Based on the  metallicity distribution in the \NHI-\EW\ plane (Fig.~\ref{fig:intro}),
we postulate that intervening absorbers in the metal rich region have one physical origin (out-flowing material),
whereas absorbers in the metal poor region are `classical DLAs', in the sense that the sight lines 
pass through the ISM of galaxies.  This simple picture naturally explains the bimodality of DLAs
reported by Wolfe et al. (see text). \label{fig:model}}
\end{figure}

% as a metallicity `ceiling' in the light of our results.

As we already noted, the \NHI\ distribution is bimodal in the \NHI-\EW\ plane (Fig.~\ref{fig:intro}a).
This is particularly true for systems with \EW$<1.5$~\AA.
This indicates there  might be two classes of intervening absorbers, probing
 different physical conditions.
While observational bias may play a role here,  
 the survey of \citet{RaoS_06a} is unbiased in regards to \HI\ column density, and
the bimodality is already present in their sample.

Given the metallicity gradient shown in the previous section and the absorber distribution in
the \NHI-\EW\ plane, we make the following assumption to guide our understanding: 
    absorbers in the red metal-rich shaded region in Fig.~\ref{fig:gradient} originate 
  in one physical environment with a more homogeneous metallicity distribution,
   while the blue metal-poor absorber 
  in the blue region with low \EW and high-\NHI\ may be the `classical'
DLAs where the sight-line is probing the ISM of small galaxies.
  These two hypothesis are illustrated with Fig.~\ref{fig:model}.

Since our recent results \citep{BoucheN_06c,BoucheN_07a}   favor  the outflow scenario
for \MgII-selected absorbers, we postulate that the homogeneous metallicity distribution originates
from the  metal-rich material  being driven out of 
sub-L$^*$ galaxies. Indeed, several studies  \citep{BoucheN_06b,OppenheimerB_08a} have shown
that galaxies with $L<1/3L^*$ dominate the metal budget in the intergalactic medium, i.e.
the metals outside the ISM of galaxies.
This material is either directly entrained by ram pressure from the hot outflow, or 
traces cooling material (where the metallicity is higher and the cooling time is shorter) 
\citep{MallerA_04a}.
Either way it may very well return to the ISM of the galaxy as recent wind models
suggest \citep{OppenheimerB_08a}.

This picture outlined in Fig.~\ref{fig:model}, which may still be somewhat over simplified, 
naturally explains many other observables of intervening
absorbers. For instance, as pointed out in the previous section, a sample of DLAs will be made 
of a mix of the two types of absorbers. 
Interestingly, \citet{WolfeA_08a} reported  evidence for a bimodality in DLAs
using [C II] 158$\mu$m cooling rates, $l_c$. 
The `low-cool' DLAs  have lower velocity widths, lower metallicity and lower dust-to-gas ratios
than the 'high-cool' DLAs which have larger velocity widths and higher metallicities.
Fig.~\ref{fig:gradient} naturally shows that DLAs with large velocity dispersions 
(as measured by \EW) would be more metal rich, than those
with low velocity widths. 
Furthermore, the UDF results of \citet{WolfeA_06a} imply that  in situ star formation 
 can be the dominant heating mechanism for the `low-cool' population only,
 consistent with our interpretation shown in Fig.~\ref{fig:model}.
We have shown that the DLA bimodality originates from the biased selection at constant \NHI.

The picture outlined in Fig.~\ref{fig:model}  
 predicts an increasing reddening $E(B-V)$ (owing to the \Z\ increase) with equivalent width \EW\ as reported by 
\citet{YorkD_06a}  \citep[see also][]{MenardB_08a}
  assuming a constant dust-to-gas ratio for \MgII\ absorbers since globally \EW\ and \NHI\ are correlated.
This assumption may be the case under the common physical nature (out-flowing material) of strong \MgII\ samples.
%Therefore, an increase reddening with $\log \NHI$  (or with \EW) as observed.
After this work was being completed, \citet{MenardB_08b} showed that \MgII\ absorbers on the \
 \NHI-\EW\ sequence shown in Fig.~\ref{fig:intro}(a) have indeed a constant  dust-to-gas ratio
Interestingly, they concluded that the dust-to-gas ratio was not consistent with that of dwarf galaxies (SMC),
therefore rejecting the alternative hypothesis often invoked for absorbers, namely that the sight-lines go through
 dwarfs near more normal galaxies. 
We note that the dust-to-gas ratio for
the data points in the upper left of Fig.~\ref{fig:intro}(a) (i.e. not on \NHI-\EW\ sequence)
 must be smaller  \citep{MenardB_08b}, as concluded by \citet{WolfeA_08a} for DLAs with low velocity widths.

\subsection{Caveats and possible systematics}

Both \citet{ProchaskaJ_08a} and \citet{LedouxC_06a} used a sample of DLAs 
at higher redshifts (with $1.7<z_{\rm abs}<4.3$) than the Rao et al. sample of low-$z$ absorbers
dominating our study.
However, Fig.~\ref{fig:intro}(b) does not change qualitatively for absorbers
with $z_{\rm abs}>1.6$ in our sample. The overall metallicity is lower,
reflecting the redshift evolution of metallicity in DLAs and sub-DLAs 
\citep{ProchaskaJ_03b,PerouxC_07a,KulkarniV_07a}.

Another difference between our analysis and \citet{LedouxC_06a} and \citet{ProchaskaJ_08a},
 is that we used \EW\ as a proxy for 
velocity width $\Delta v$ \citep{EllisonS_06a} whereas they used the measured velocity width from high-resolution
spectra of the low-ions \SiII\ and \ZnII. However, for the dozen of absorbers with both \EW\ and
 $\Delta v(\SiII$~or~$\ZnII)$, we find that the $\Delta v$(\EW) correlates with  $\Delta v(\SiII$~or~$\ZnII)$
 at 99\%\ confidence level. Thus, the selection effect shown in Fig.~\ref{fig:gradient}
 exists against \EW\ \citep[as in][]{MurphyM_07a} and against $\Delta v(\SiII$~or~$\ZnII)$
 \citep[as in][]{LedouxC_06a,ProchaskaJ_08a}.

One might invoke dust obscuration to account for the results shown in Fig.~\ref{fig:gradient}.
However, such a dust-bias would have to selectively remove metal rich DLAs with low \EW.
The results of \citet{YorkD_06a} and \citet{MenardB_08a}
 showed that dust obscuration in \MgII-selected samples with
the lowest \EW\ is   low, i.e. $E(B-V)<0.02$ below $\EW<2$~\AA, which
 implies a very low fraction of missed absorbers below $\EW<2$~\AA\ irrespective of \NHI.
Similarly, such a dust-bias would have to selectively remove metal poor DLAs with high \EW.
This is unlikely since dusty  and metal poor (\Z$ < $-1.0) DLAs 
would have to have unphysically high dust-to-metal ratios (${\cal D}\ge 10\cal D_{\rm MW}$),
 which is rather difficult to produce   \citep{InoueA_03a}.
 On the other hand, metal-rich DLAs with high \EW\ would have larger $E(B-V)$ \citep{MenardB_08b} and therefore
 would be easier to obscure, but are in fact present in our sample.
Finally, we looked at the distribution of the QSO magnitudes in the \NHI-\EW\ plane,
and found no evidence for a selective dust-bias. 

\section{Conclusions}

Our results show  the presence of a metallicity gradient in the \NHI-\EW\ plane
of intervening absorbers (Fig.~\ref{fig:intro}). In other words, the metallicity
of absorbers is a bimodal function of \NHI\ and \EW.
As a direct consequence,  a population of DLAs, selected with $\log \NHI>20.3$, will
   be heterogeneous. At low \EW, the \HI-selected sample is metal poor,
whereas at high \EW, it is found to be more metal rich.
Therefore, the correlation between the  metallicity
\Z\ and  the line-of-sight velocity width $\Delta v$ reported by \citet{LedouxC_06a}
and by \citet{MurphyM_07a} arises
from the \HI\ selection and can not be interpreted 
as a signature of the mass-metallicity relation akin to normal field galaxies. 

We argue that the bivariate distribution  [\Z(\NHI,\EW)] can be explained
by two different physical origins of absorbers, which are the ISM of small galaxies and
  out-flowing material, with distinct physical properties (such as metallicities and dust-to-gas ratios).
This is supported by the  distribution of the absorbers in the \NHI-\EW\ plane.
If there are two distinct populations of absorbers, as shown in Fig.~\ref{fig:model},
this naturally explains
the two classes of DLAs (`low-cool' and `high-cool'), reported by \citet{WolfeA_08a} using [C II] 158$\mu$m cooling rates.
 DLAs with large velocity dispersions 
(as measured by \EW) are more metal rich  than those
with low velocity widths and will have different dust-to-gas ratio for a given dust-to-metal ratio.

Therefore, the correlation between metallicity \Z\ and \EW(or $\Delta v$) for DLAs (which we showed to be apparent),
the two classes of DLAs of \citet{WolfeA_08a}, the dust-to-gas results of \citet{MenardB_08b},
and the results of \citet{BoucheN_06c} indicating that \MgII-selected absorbers
are tracing out-flowing material can all be put into one coherent context.

\section*{Acknowledgments}
We thank S. Ellison for providing  \EW\ for her sample.
We acknowledge inlighting scientific discussions with M.T. Muprhy, C. P\'eroux.
We gratefully acknowledge   M.T. Murphy and S. Genel for their thorough reading
of the manuscript.

%\bibliography{references}
%\bibliographystyle{mn2e}

\bsp_small

\label{lastpage}

\end{document}